\documentclass[12pt]{article}
\usepackage{amsmath,amssymb,amsfonts,amsbsy}
\usepackage{cite}
\usepackage{epsfig}
\usepackage{graphicx}
\usepackage{wrapfig}
\usepackage[font=small,labelfont=bf,labelsep=period]{caption}

\newcommand{\be}{\begin{equation}}
\newcommand{\ee}{\end{equation}}
\newcommand{\ba}{\begin{eqnarray}}
\newcommand{\ea}{\end{eqnarray}}

\newcommand{\tvec}[1]{\mbox{\boldmath{$#1$}}}

\begin{document}
\addcontentsline{toc}{subsection}{{Transverse momentum dependent parton distributions and azimuthal asymmetries in light-cone quark models}}
{\it }
\setcounter{section}{0}
\setcounter{subsection}{0}
\setcounter{equation}{0}
\setcounter{figure}{0}
\setcounter{footnote}{0}
\setcounter{table}{0}
\begin{center}
\textbf{TRANSVERSE MOMENTUM DEPENDENT PARTON DISTRIBUTIONS AND AZIMUTHAL ASYMMETRIES IN LIGHT-CONE QUARK MODELS}\\
\vspace{5mm}
B. Pasquini$^{\,1\,\dag}$, S. Boffi,$^{\,1}$, A.~V.~Efremov$^{\, 2}$, 
and P. Schweitzer$^{\,3}$\\
\vspace{5mm}
\begin{small}
  (1) \emph{
Dipartimento di Fisica Nucleare e Teorica, Universit\`{a} degli Studi di 
Pavia, and \\ Istituto Nazionale di Fisica Nucleare, Sezione di Pavia, I-27100 
Pavia, Italy}\\
  (2) \emph{
Joint Institute for Nuclear Research, Dubna,141980 Russia} \\
  (3) 
\emph{Department of Physics, University of Connecticut, Storrs, CT 06269, USA} 
\\
  $\dag$ \emph{pasquini@pv.infn.it}
\end{small}
\end{center}
\vspace{0.0mm} 
\begin{abstract}
We review the information  on the spin and orbital angular momentum structure
of the nucleon encoded in the T-even 
transverse momentum dependent parton distributions within light-cone quark models. 
Model results for azimuthal spin asymmetries in semi-inclusive 
lepton-nucleon  deep-inelastic scattering  are discussed, showing a good 
agreement with available experimental data and providing
predictions to be further tested by future CLAS, COMPASS and HERMES data.
\end{abstract}
\vspace{7.2mm} 

\section{TMDs and Light-Cone CQMs}

A convenient framework for the analysis of hadronic states is quantization 
on the light-cone.
The proton state, for example, can be represented as a superposition of light-cone wave functions (LCWFs), 
one for each of the Fock components $(qqq)$, $(qqq\bar q)$, ... of the nucleon state.
This light-cone representation  has a number of simplifying properties\cite{BPP98}.
In particular it allows one to describe the hadronic matrix elements which parametrize
the soft-contribution in inclusive and exclusive reactions in terms of overlap of LCWFs with different parton configurations.
In principle, there is an infinite number of LCWFs in such an expansion.
However, there are many situations where one can confine the analysis to the contribution of the 
Fock components with a few partons.
For example, light-cone  models limited to the minimal
 Fock-space configuration of valence quarks are able to reproduce the  main features of the hadron 
electromagnetic 
form factors~\cite{Pasquini:2007iz} as well to account for the behaviour of the hadron structure functions in
deeply inelastic processes at large values of the 
Bjorken variable $x$~\cite{Boffi:2007yc,Pasquini:2005dk}.
\newline
\noindent
Here the LCWFs of constituent quark models (CQMs) will be used to describe transverse momentum dependent parton distributions (TMDs)
which are a natural extension of standard parton distributions from one to three-dimensions in momentum space. 
In particular, to disentangle the spin-spin and spin-orbit quark correlations encoded in the different TMDs, we expand the three-quark LCWF in a basis of eigenstates of orbital angular momentum.
In the light-cone gauge $A^+=0$, such an expansion involves six independent amplitudes corresponding to
the different combinations of quark helicity 
and 
orbital angular momentum. 
Explicit expressions for the light-cone amplitudes have been obtained in Ref.~\cite{Pasquini:2008ax}
representing the light-cone spinors of the quarks through the unitary Melosh rotations which 
 boost the rest-frame spin into the light-cone.
Furthermore, assuming SU(6) symmetry, the light-cone amplitudes have a particularly simple structure,
with the spin and isospin dependence factorized from a momentum-dependent function which is 
spherically symmetric.
Under this assumption the orbital angular momentum content 
of the wave function is fully generated by the Melosh rotations and therefore matches
the analytical structure expected from model-independent arguments\cite{Ji:2002xn}.
The model dependence enters  the choice of the momentum-dependent part of the LCWF,
as obtained, for example, from
the eigenvalue equation of the Hamiltonian with a specific potential 
model for the bound state of the three quarks.
Using a more phenomenological description, we choose this part by assuming a specific functional form with parameters fitted to hadronic structure constants.
This is the strategy adopted also in Ref.\cite{Schlumpf:1992ce} through a fit of the LCWF
to the anomalous magnetic moments of the nucleon.
The same wave function was also used to predict many other hadronic 
properties~\cite{Schlumpf:1992pp},
providing a good description of available experimental data, and being able
to capture
the main features of hadronic structure functions,
like parton distributions~\cite{Pasquini:2005dk},  
generalized parton distributions (GPDs)~\cite{Boffi:2007yc} and TMDs\cite{Pasquini:2008ax}.
\newline
\noindent
The eight leading twist TMDs, 
$f_1,$ $f_{1T}^{\perp },$ $g_1,$  $g_{1T},$ $g_{1L}^{\perp},$
$h_1,$  $h_{1T}^{\perp},$ $h_{1L}^{\perp}$, and 
$h_{1}^{\perp},$
are defined in terms of the same quark correlation functions entering
the definition of ordinary parton distributions, but without integration over the transverse momentum.
 Among them,
 the Boer-Mulders $h_1^\perp$\cite{Boer:1997nt} and the Sivers $f_{1T}^\perp$\cite{Sivers:1989cc} functions are T-odd, 
i.e. they change sign under ``naive time reversal'', 
which is defined as usual time reversal, but without interchange of initial 
and final states.
Since non-vanishing T-odd TMDs require gauge boson degrees of freedom which are not taken into account in our light-cone quark model, 
our model results will be discussed only for the T-even TMDs.
\newline
\noindent
Projecting the correlator for quarks of definite longitudinal ($s_L$) or 
transverse ($\tvec{s}_T$) polarizations, one obtains in nucleon states
described by the polarization vector $\tvec{S}=(S_L,\tvec{S}_T)$
the following spin densities in the momentum space
\begin{eqnarray}
\label{piet-distr1}
\tilde\rho(x,\tvec{k}_T^2, s_L,\tvec{S})
 &=& \frac{1}{2} \Bigg[ f_1^{\phantom{\perp}\!\!}
   + S_T^i \epsilon^{ij} k_T^j \frac{1}{m}\, f_{1T}^\perp
   +  s_L  S_L\, g_{1L}^{\phantom{\perp\!\!}}
   +  s_L\, S_T^i k_T^i \frac{1}{m}\, g_{1T}^{\phantom{\perp\!\!}} 
  \Bigg] \, ,
\\
\tilde\rho(x,\tvec{k}_T^2,\tvec{s}_T,\tvec{S}_T)
 &=& \frac{1}{2} \Bigg[ f_1^{\phantom{\perp\!\!}}
   + S^i_T \epsilon^{ij} k_T^j \frac{1}{m}\, f_{1T}^\perp
   + s^i_T \epsilon^{ij} k_T^j \frac{1}{m}\, h_{1}^\perp
   + s^i_T S^i_T h_1^{\phantom{\perp\!\!}}
\nonumber \\
  \label{piet-distr2}
 && \hspace{0.7em}
 {}+ s^i_T (2 k_T^i k_T^j - \tvec{k}_T^2 \delta^{ij}) S^j_T 
       \frac{1}{2m^2}\, h_{1T}^\perp
   +   S_L\, s^i_T k_T^i \frac{1}{m}\, h_{1L}^\perp \Bigg] \, ,
\end{eqnarray}
where the distribution functions depend on $x$ and $\tvec{k}_T^2$.
The unpolarized TMD $f_1$, the helicity TMD $g_{1L}$, and the transversity TMD $h_1$ in Eqs.~(\ref{piet-distr1}) and (\ref{piet-distr2}) correspond to 
monopole distributions in the momentum space for unpolarized, longitudinally and transversely polarized nucleon, respectively.
They can be obtained from the overlap of LCWFs which are diagonal in the 
orbital angular momentum, but probe different transverse momentum and helicity correlations of the quarks inside the nucleon.
All the other TMDs require a transfer of orbital angular momentum
between the initial and final state.
In particular,  $g_{1T}$  and $h_{1L}^\perp$ correspond to  quark densities
with specular configurations for the quark and nucleon spin:
$g_{1T}$ describes 
longitudinally polarized quarks in a transversely polarized nucleon, while
 $h_{1L}^\perp$ gives the distribution of transversely polarized quarks in longitudinally polarized 
nucleon.
Therefore, $g_{1T}$ requires helicity flip of the nucleon
which is not compensated by a change of the quark helicity, and viceversa $h_{1L}^\perp$ involves helicity flip of the quarks 
but is diagonal in the nucleon helicity. 
As a result, in both cases, the LCWFs of the initial and final states
differ by one unit of orbital angular momentum and the associated spin distributions have a dipole structure.
Finally, for transverse polarizations in perpendicular directions
 of both the quarks and the nucleon, one has a quadrupole distribution with strength 
given by  $h_{1T}^\perp$. In this case, the nucleon helicity flips in the direction 
 opposite to the quark helicity, with a mismatch of two units 
for the orbital angular 
momentum of the initial and final LCWFs.

\begin{figure}[t]
\centerline{\psfig{file=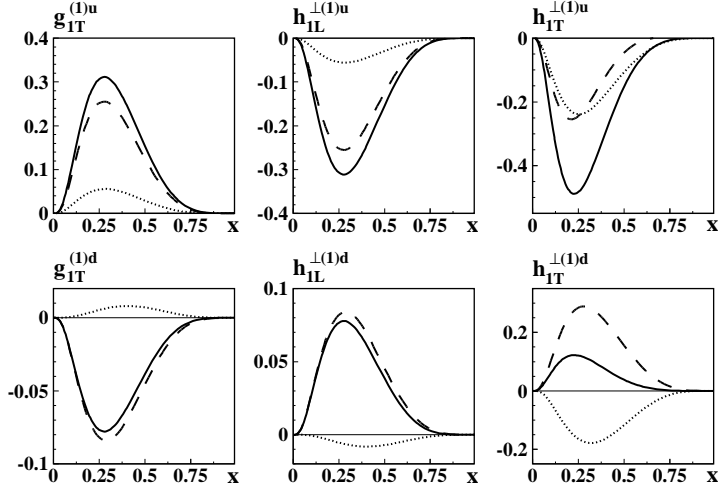, width=10 cm}}
\vspace*{8pt}
\caption{Transverse moments of 
TMDs
as function of $x$ for up (upper panels) and down (lower panels) quark.
The solid curves show the total results, 
sum of the partial wave contributions.
In the case of $ g_{1T}^{(1)}$ and $h_{1L}^{\perp(1)}$ the dashed and dotted curves give the results from the S-P and P-D interference terms, respectively. 
In the case of $h_{1T}^{\perp(1)}$, the dashed curve  is the result from P-wave interference, and the dotted curve is due to the interference of S and D waves.
\protect\label{fig3}}
\end{figure}

\begin{figure}[t]
\centerline{\hspace{-1 cm}\psfig{file=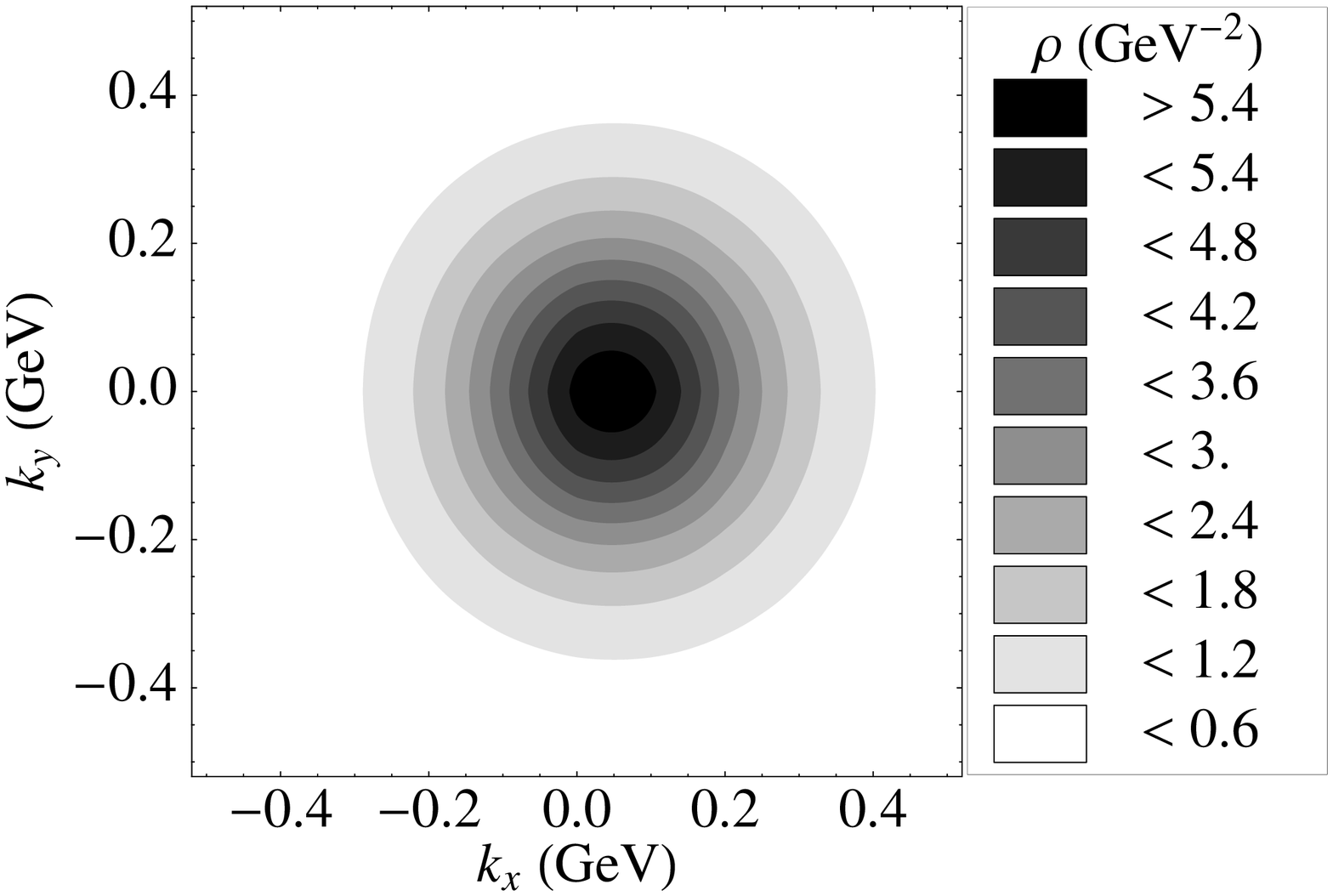, width=5.5 cm}
\hspace{0.5 cm}
\psfig{file=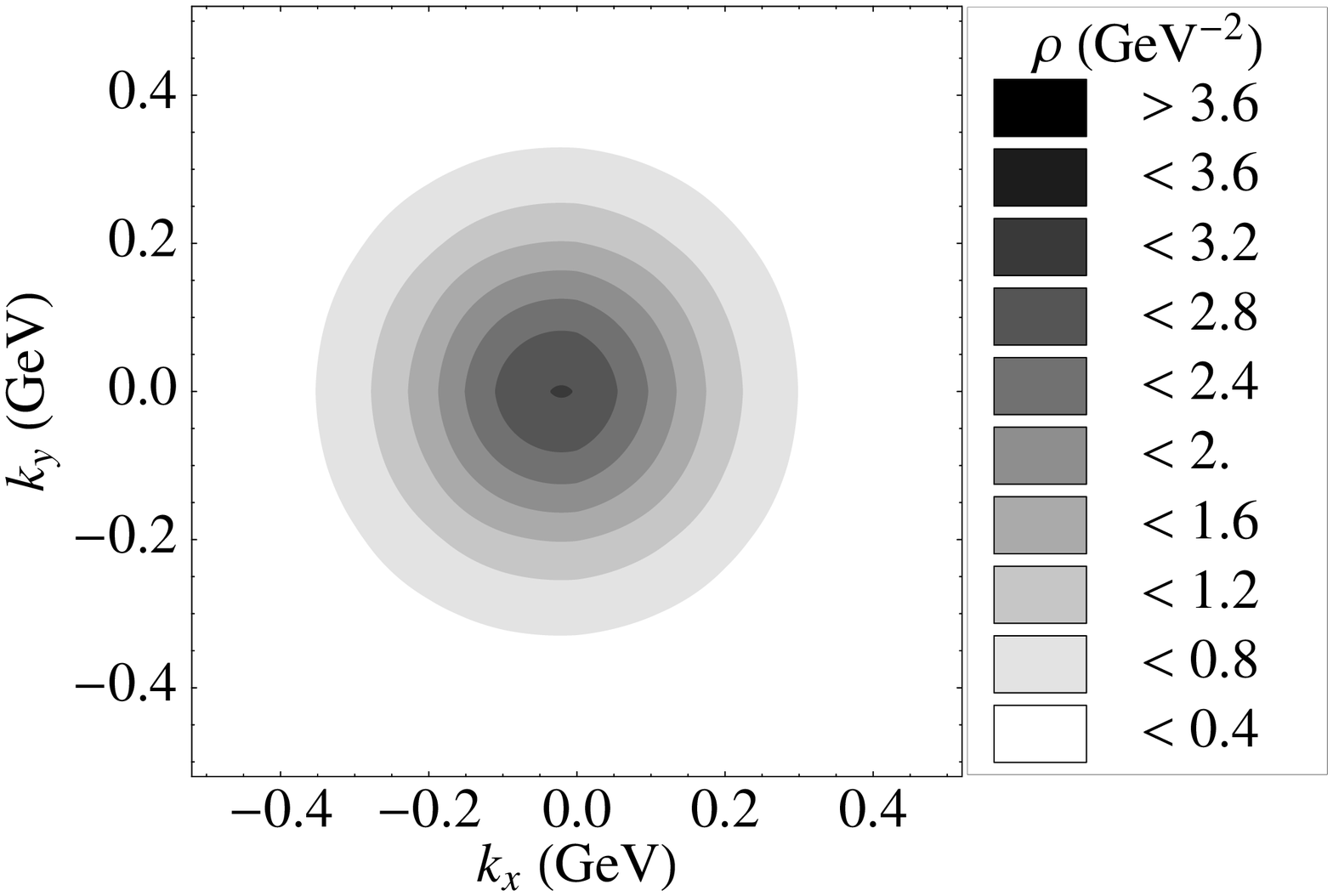, width=5.5 cm}}
\vspace{-20pt}
\caption{Quark densities in the $\tvec{k}_T$ plane
for longitudinally polarized quarks in a transversely polarized 
proton for up (left panel ) and down (right panel) quark.
\protect\label{fig2}}
\end{figure}

In Fig.~\ref{fig3} is shown the interplay between the different 
partial-wave contributions to the transverse moments  $ g_{1T}^{(1)}$, $h_{1L}^{\perp(1)}$ and $h_{1T}^{\perp(1)}$, defined as 
$g_{1T}^{(1)}(x)=\int{\rm d}^2\tvec{k}_T(\tvec{k}^2_T/2m^2)g_{1T}^{\phantom{\perp\!\!}}(x,\tvec{k}^2_T)$, etc.
While the first two functions $ g_{1T}^{(1)}$ and $h_{1L}^{\perp(1)}$ are dominated by the
 contribution due to P-wave interference, in the case of $h_{1T}^{\perp(1)}$
the contribution from the D wave is amplified through the interference 
with the S wave. The total results for up and down quarks
obey the SU(6) isospin relation, i.e. the functions for up quarks 
are four times larger than for down quark and with opposite sign.
This does not apply to the partial-wave contributions, as it is evident in 
particular for the terms containing D-wave contributions.
\newline
\noindent
Among the distributions in Eqs.~(\ref{piet-distr1}) and  (\ref{piet-distr2}), the dipole 
correlations related to $g_{1T}$ and $h_{1L}^{\perp}$ have characteristic features
of intrinsic transverse momentum, since they are the only ones which have  no
analog in the spin densities
related to the GPDs in the impact parameter space~\cite{Diehl:2005jf,Pasquini:2007xz}.
The results in the light-cone quark model of Ref.\cite{Pasquini:2008ax} 
for the densities 
with longitudinally polarized quarks in a transversely polarized 
proton are shown in Fig.~\ref{fig2}.
The sideways shift in the positive (negative) $x$ direction for up (down) quark 
due to the dipole term $\propto  s_L\, S_T^i k_T^i \frac{1}{m}\, g_{1T}^{\phantom{\perp\!\!}}$
is sizable, and corresponds to an average deformation
$\langle \tvec{k}_x^u\rangle=55.8 $ MeV, and  
$\langle \tvec{k}_x^d\rangle=-27.9 $ MeV. 
The dipole distortion  
$\propto S_L\, s_T^i k_T^i \frac{1}{m}\, h_{1L}^\perp$
in the case of transversely polarized quarks in a longitudinally polarized proton is equal but with opposite sign, 
since in our model $h_{1L}^\perp=-g_{1T}$.
(Also other quark model relations among TMDs \cite{Avakian:2008dz} 
    are satisfied in our model, see \cite{Pasquini:2008ax}.)
These model results are supported from a recent lattice 
calculation\cite{Hagler:2009mb,Musch:2009ku} 
which gives, 
for the density related to $g_{1T}$,
$\langle \tvec{k}_x^u\rangle=67(5) $ MeV, and  
$\langle \tvec{k}_x^d\rangle=-30(5) $ MeV. For the density
related to  $h_{1L}^\perp$,  they also find shifts of similar magnitude but opposite sign:
$\langle \tvec{k}_x^u\rangle=-60(5) $ MeV, and  
$\langle \tvec{k}_x^d\rangle=15(5) $ MeV.

\section{Results for azimuthal SSAs}

\begin{figure}[b]
\centerline{
\psfig{file=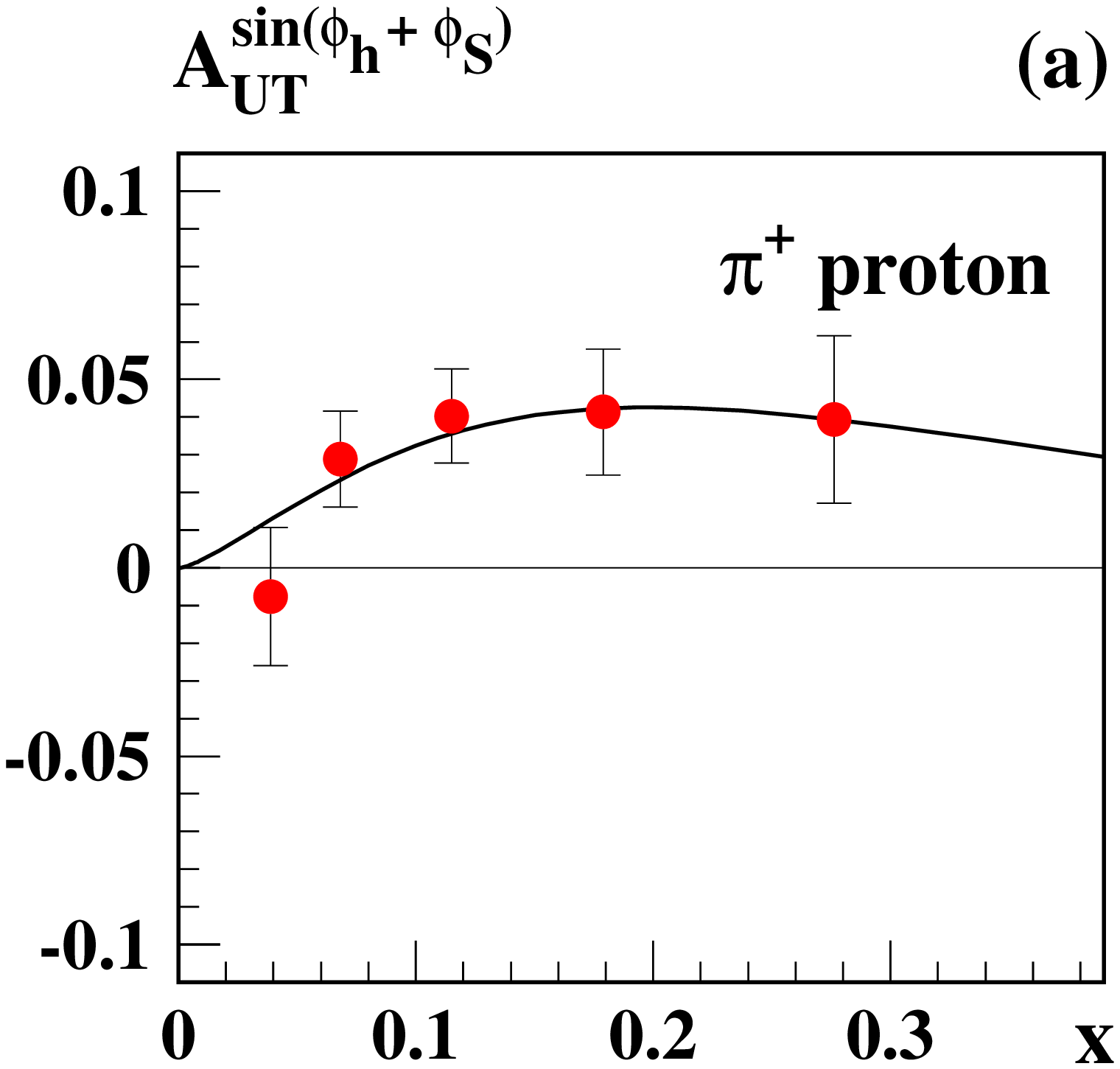, height=4.6 cm}
 \hspace{-11mm}
\psfig{file=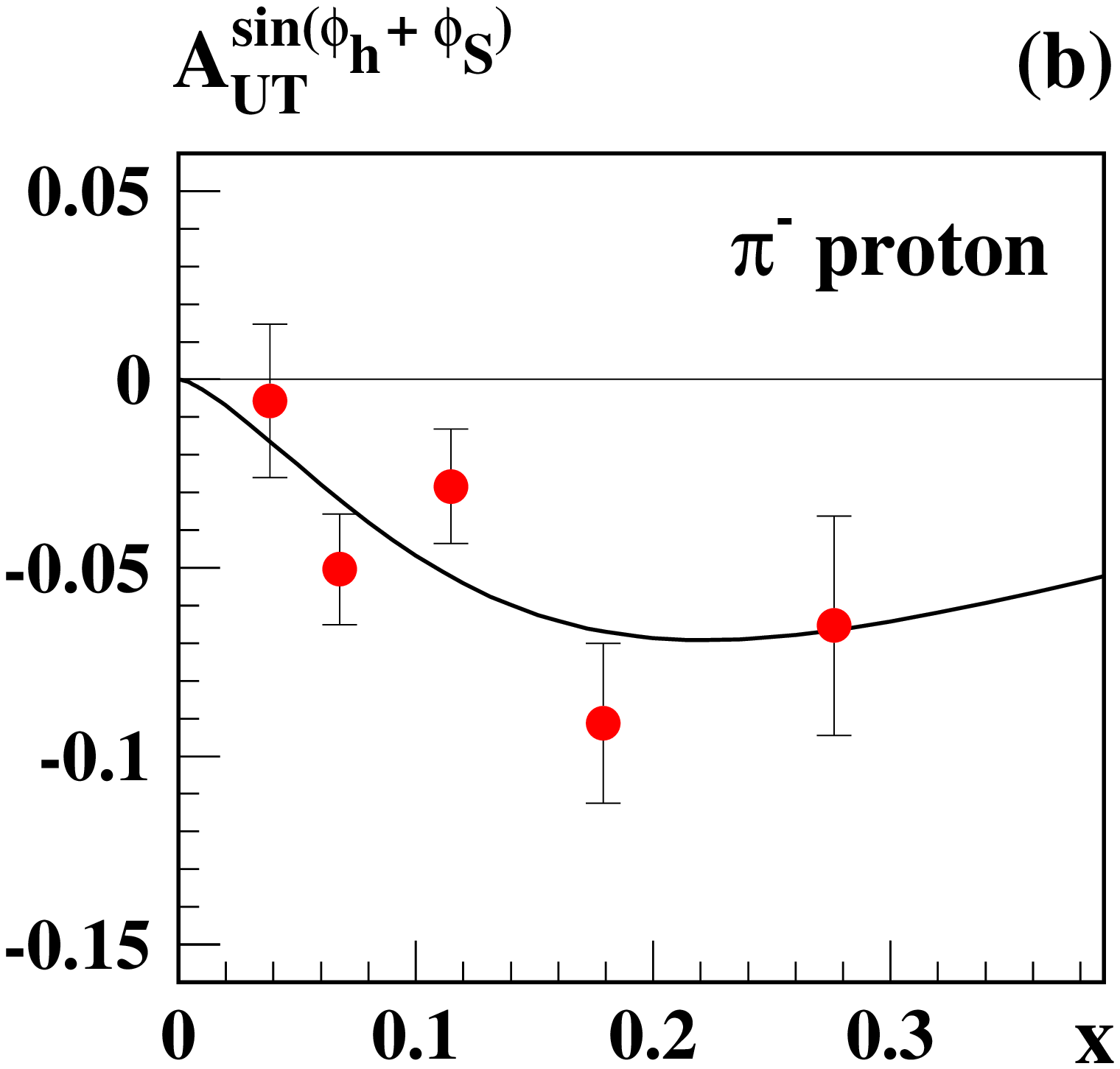, height=4.6 cm}
 \hspace{-12mm}
\psfig{file=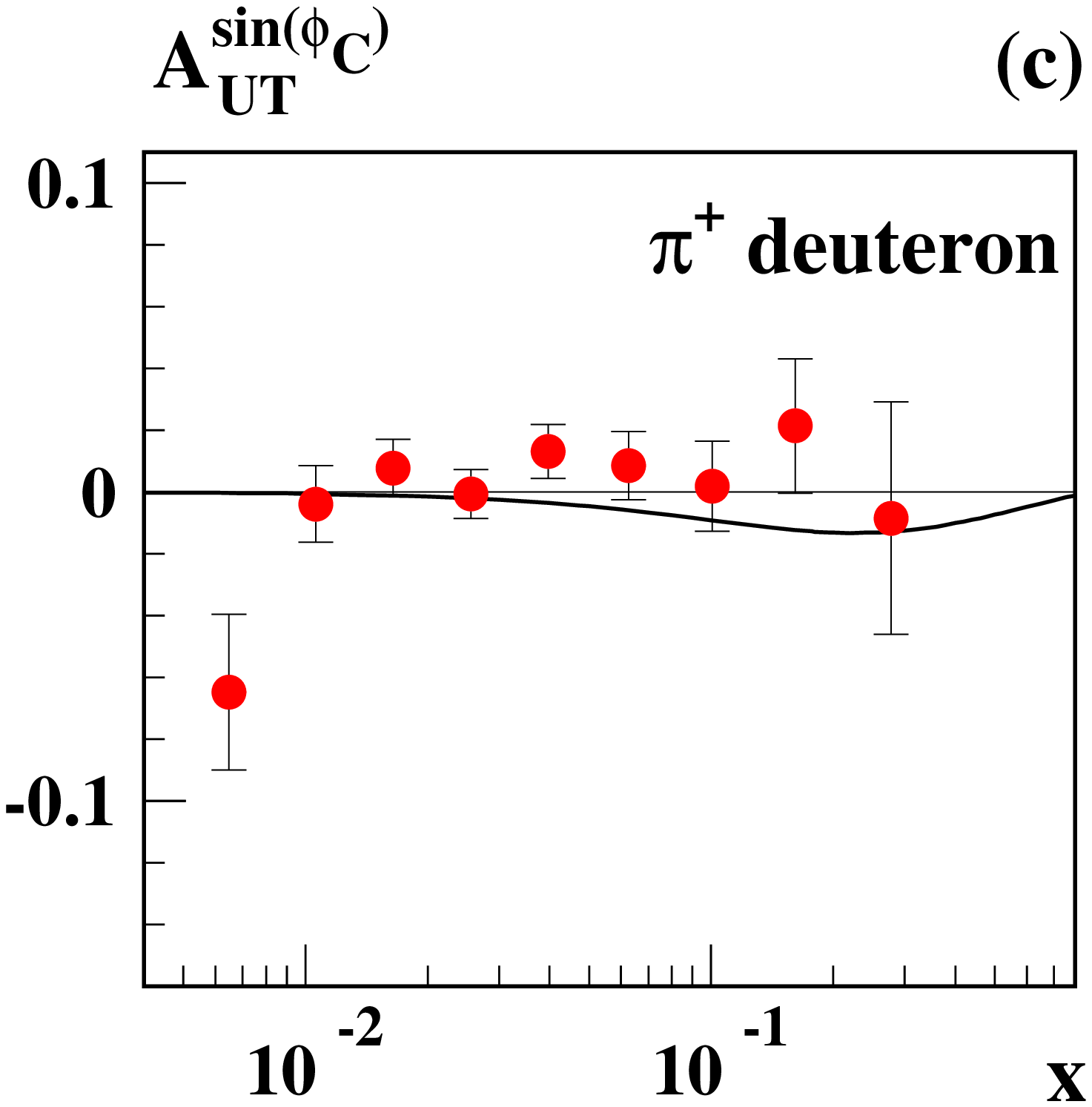, height=4.6 cm}
 \hspace{-11mm}
\psfig{file=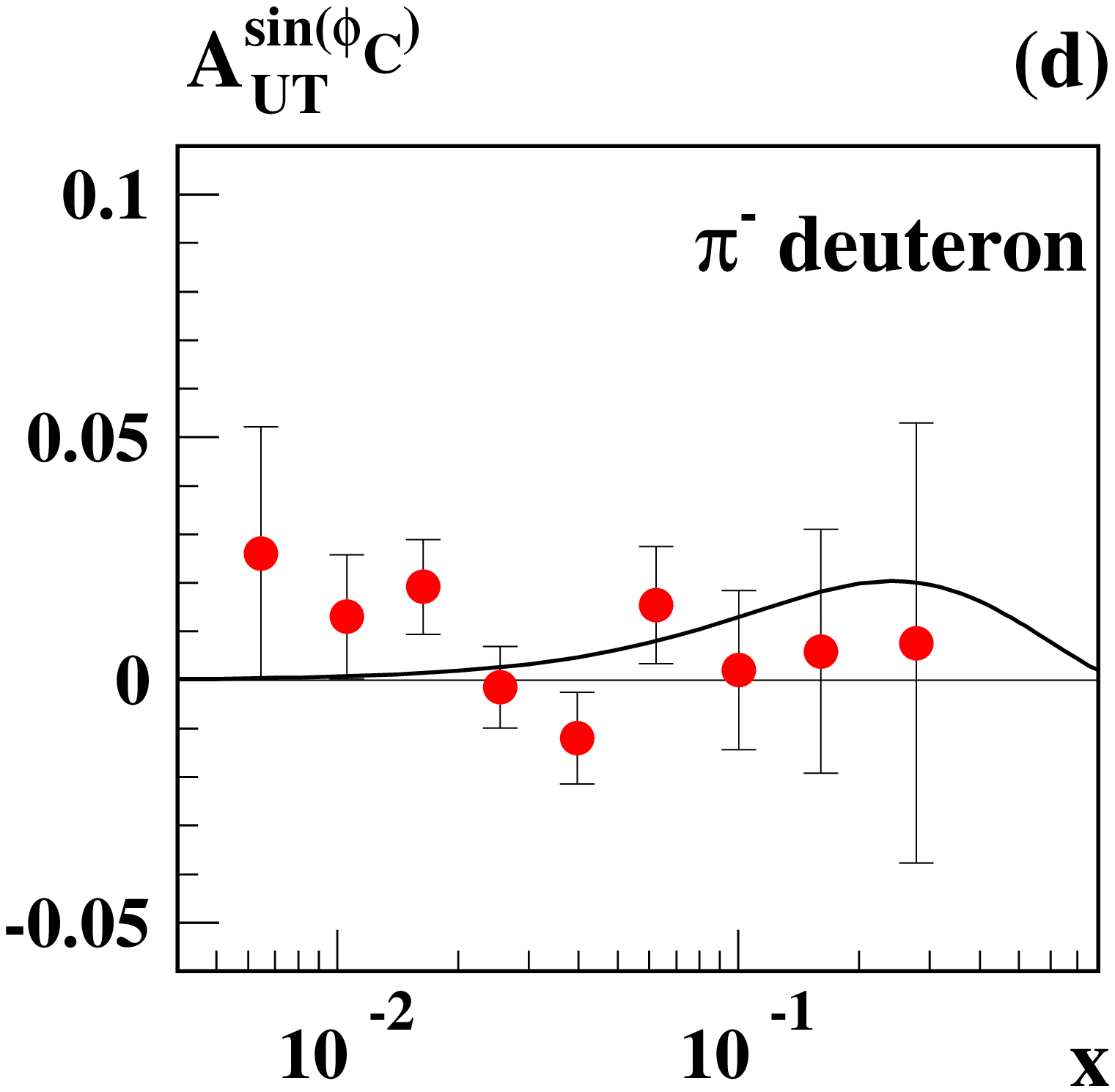, height=4.6 cm}
}
	\caption{\label{fig4}
	The single-spin asymmetry 
        $A_{UT}^{\sin(\phi_h+\phi_S)}\equiv-A_{UT}^{\sin\phi_C}$ in DIS
	production of charged pions 
	off proton and deuterium targets, as function 
        of $x$. The theoretical curves are obtained on the basis of the 
        light-cone CQM predictions for $h_1(x,Q^2)$ from
	Ref.~\protect\cite{Pasquini:2005dk,Pasquini:2008ax}.
        The (preliminary) proton target data are 
        from HERMES \protect\cite{Diefenthaler:2005gx}, 
	the deuterium target data are from COMPASS \protect\cite{Alekseev:2008dn}.}
\end{figure}
\begin{figure}[t]
\vspace{1 truecm}
\centerline{
 \includegraphics[height=3.75cm]{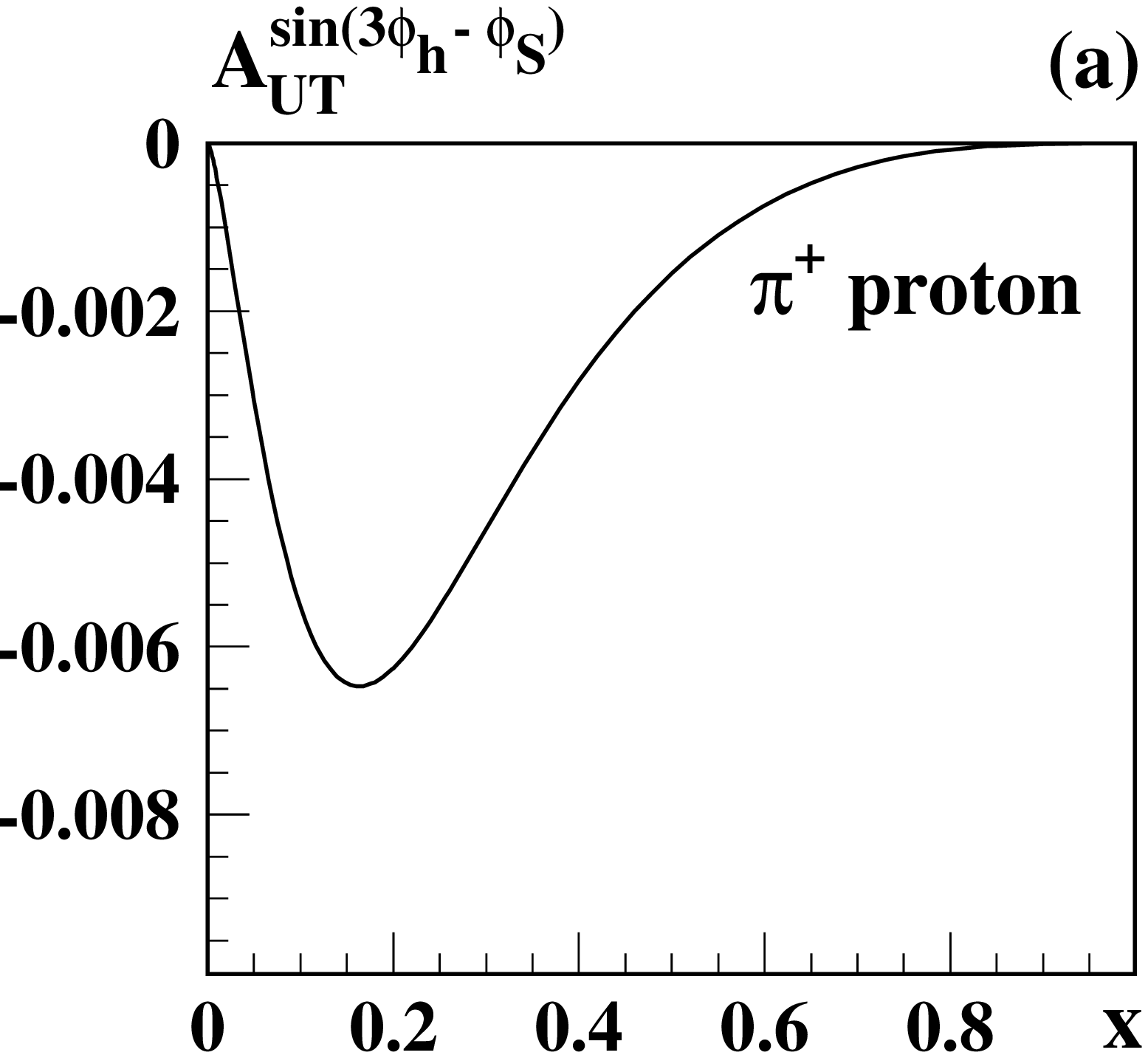}
 \hspace{-11mm}
 \includegraphics[height=3.75cm]{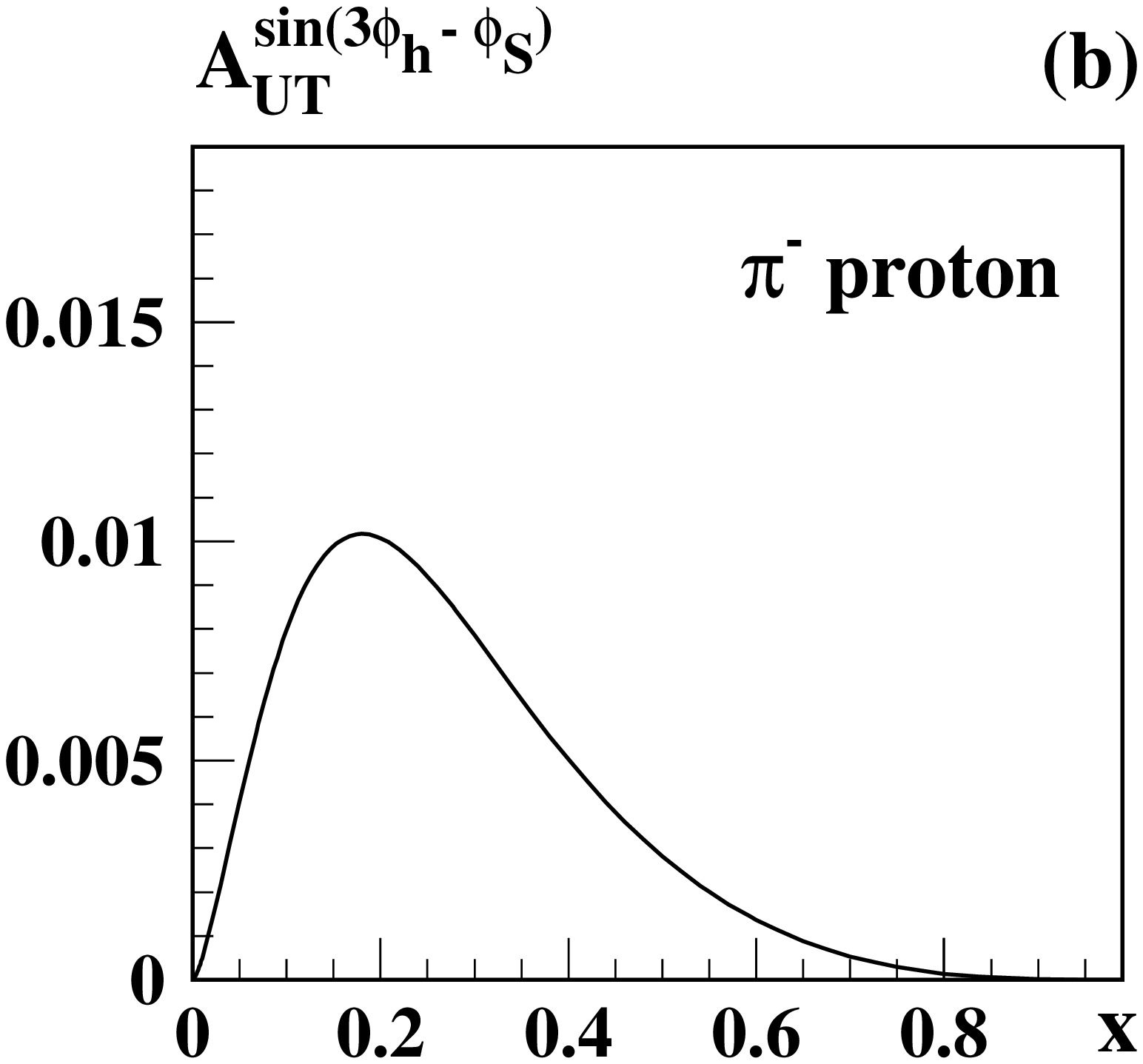}
 \hspace{-10mm}
 \includegraphics[height=3.75cm]{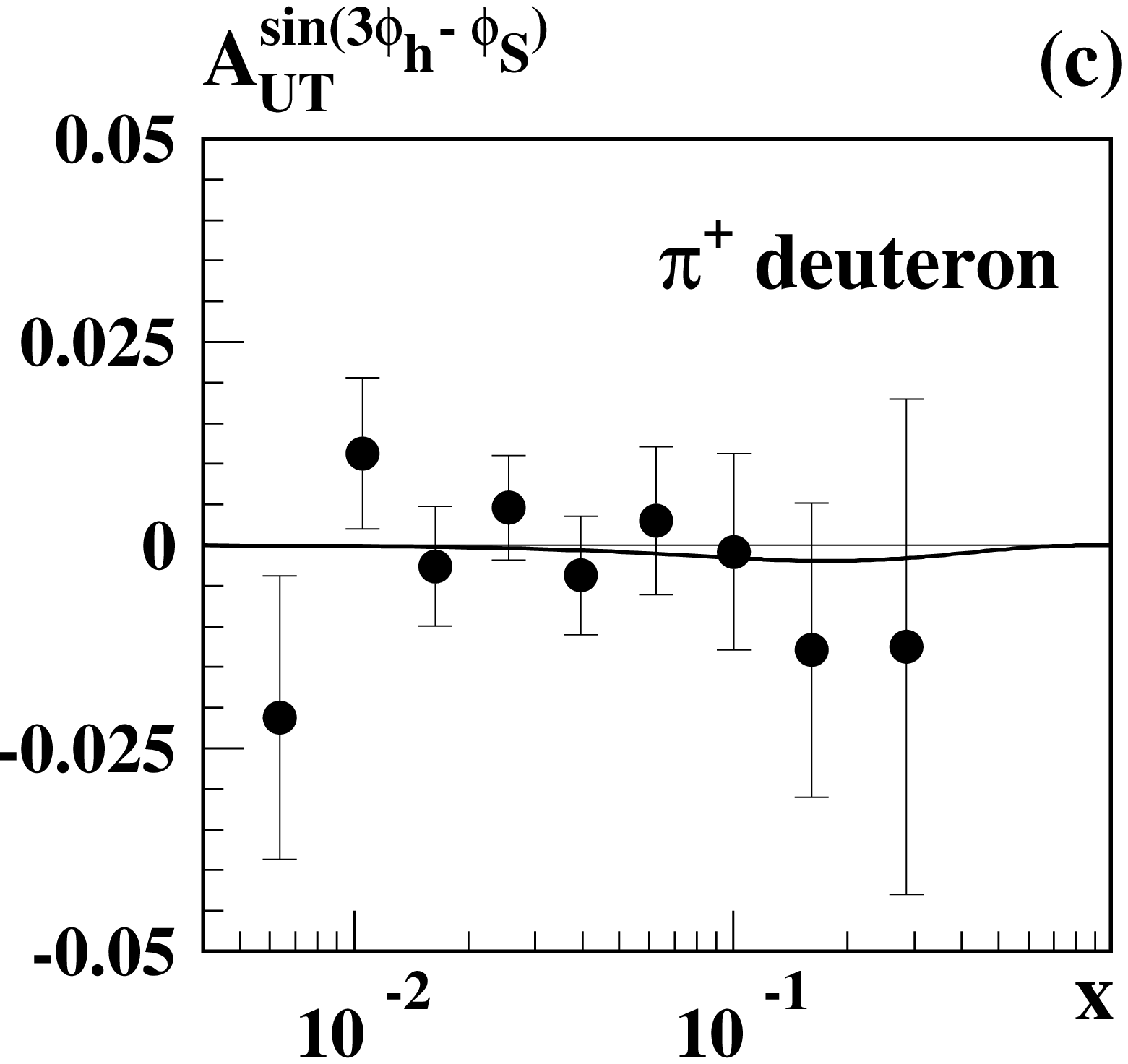}
 \hspace{-10mm}
 \includegraphics[height=3.75cm]{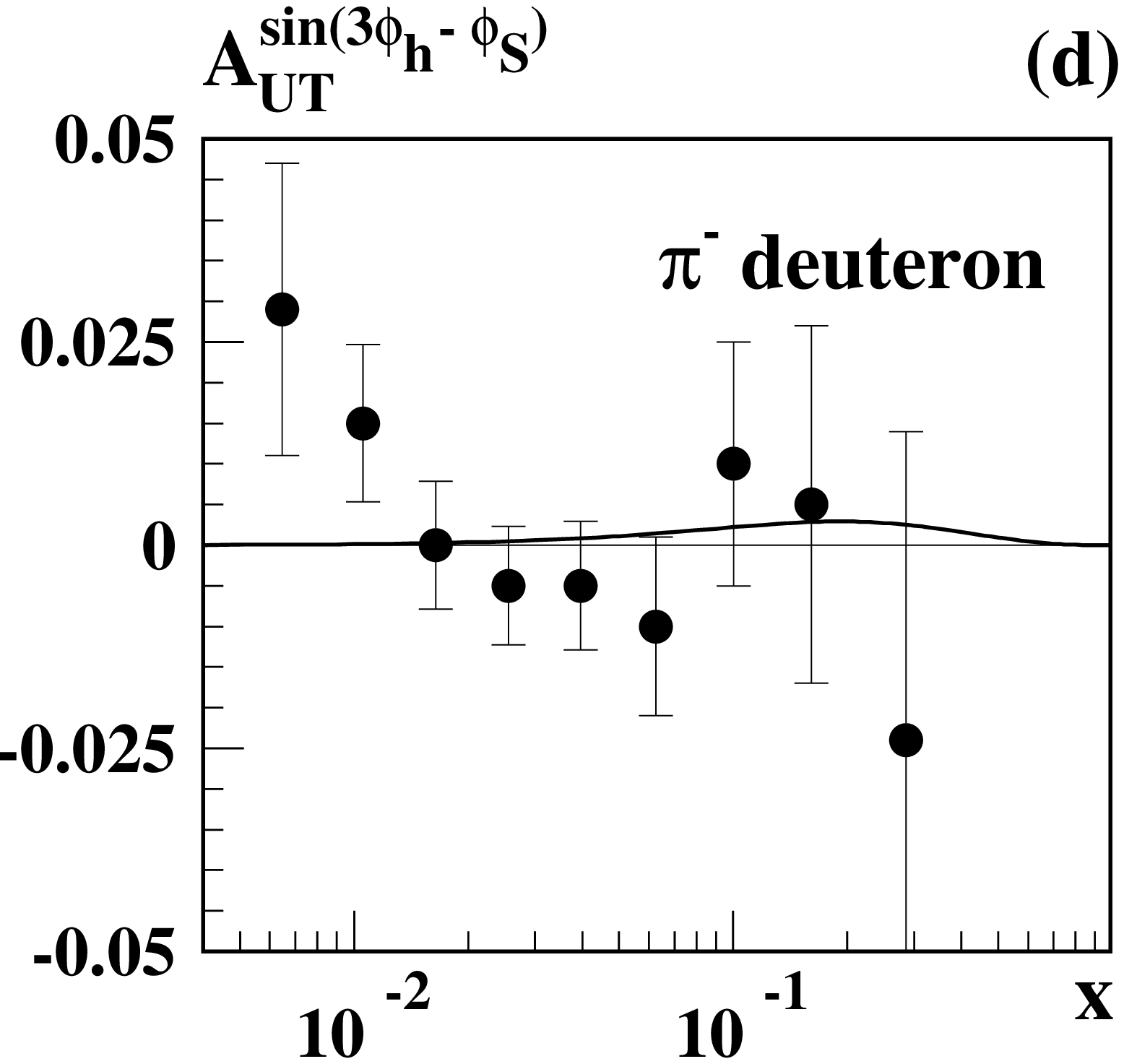}
}
{
    	\caption{\label{autbis}
	The single-spin asymmetry $A_{UT}^{\sin(3\phi_h-\phi_S)}$ in DIS
	production of charged pions off proton and deuterium targets, 
	as function of $x$.
	The theoretical curves are obtained by evolving
	the light-cone CQM predictions for $h_{1T}^{\perp(1)}$ of
	Ref.~\protect\cite{Pasquini:2008ax} to $Q^2=2.5$ GeV$^2$, using
	the $h_1$ evolution pattern.
        The preliminary COMPASS data are from Ref.~\protect\cite{Kotzinian:2007uv}.
}
}
\end{figure}

In Ref.\cite{Boffi:2009sh} the present results for the T-even TMDs 
were applied to estimate azimuthal asymmetries in SIDIS, discussing
the range of applicability of the model, especially with regard to the scale
dependence of the observables and the transverse-momentum dependence of the 
distributions.
Here we review the results  for the Collins asymmetry
 $A_{UT}^{\sin(\phi+\phi_S)}$  and for  $A_{UT}^{\sin(3\phi-\phi_S)}$, due to the Collins fragmentation function and to the chirally-odd TMDs $h_1$,  and 
$h_{1T}^\perp$, respectively.
In both cases, we use the results extracted in \cite{Efremov:2006qm}
for the Collins function.
In the denominator of the asymmetries we take $f_1$ from\cite{Gluck:1998xa}
and the unpolarized fragmentation function from\cite{Kretzer:2000yf}, both valid at the scale $Q^2=2.5$ GeV$^2$.
\newline
In Fig.~\ref{fig4} the results for the Collins asymmetry
in DIS production of charged pions off proton and deuterium targets are shown 
as function of $x$.
The model results for $h_1$ evolved from the low hadronic scale
of the model to $Q^2=2.5 $ GeV$^2$ ideally describe the HERMES 
data\cite{Diefenthaler:2005gx} for a proton target (panels (a) and (b) of Fig.~\ref{fig4}).
This is in line with the favourable comparison 
between our model predictions\cite{Pasquini:2005dk}  and the phenomenological extraction 
of the transversity and the tensor charges in Ref.\cite{Anselmino:2007fs}.
Our results are compatible also with the COMPASS data\cite{Alekseev:2008dn}
for a deuterium target (panels (c) and (d) of Fig.~\ref{fig4}) which extend down to much lower values of $x$.
\newline
In the case of the asymmetry  $A_{UT}^{\sin(3\phi-\phi_S)}$ we face the question
how to evolve $h_{1T}^{\perp(1)}$ from the low scale of the model to the relevant
experimental scale.
Since exact evolution equations are not available in this case, we ``simulate'' the evolution of  $h_{1T}^{\perp(1)}$ by evolving it according to the transversity-evolution pattern. 
Although this is not the correct evolution pattern, 
it may give us a rough insight on the possible size of effects 
due to evolution 
(for a more detailed discussion we refer to\cite{Boffi:2009sh}).
The evolution effects give 
smaller  asymmetries in absolute value and shift the peak at lower $x$ values
in comparison with the results obtained without evolution.
The results shown in Fig. \ref{autbis} are also much smaller than the 
bounds allowed by positivity, 
$|h_{1T}^{\perp(1)}|\leq\frac12\,(f_1(x)-g_1(x)) $, and
constructed using parametrizations of the unpolarized and helicity distributions at $Q^2=2.5$ GeV$^2$.
Precise measurements in  range $0.1\lesssim x \lesssim 0.6$ are planned with the CLAS 12 GeV upgrade\cite{Avakian-LOI-CLAS12} 
and will be able to discriminate between these two scenarios.
There exist also preliminary 
deuterium target data\cite{Kotzinian:2007uv}
which are compatible, within error bars,
with the model predictions both at the hadronic and the evolved scale.

\section*{Acknowledgments}
This work is part of the activity HadronPhysics2, Grant Agreement n. 227431, 
under the Seventh Framework Programme of the European Community. It
is  also supported in part by DOE contract DE-AC05-06OR23177, 
the Grants RFBR 09-02-01149 and 07-02-91557, RF
MSE RNP 2.1.1/2512 (MIREA) and by the Heisenberg-Landau Program of JINR.



\bibliographystyle{aipproc}   

\bibliography{sample}

\IfFileExists{\jobname.bbl}{}
 {\typeout{}
  \typeout{******************************************}
  \typeout{** Please run "bibtex \jobname" to optain}
  \typeout{** the bibliography and then re-run LaTeX}
  \typeout{** twice to fix the references!}
  \typeout{******************************************}
  \typeout{}
 }

\bibliographystyle{aipproc}   

\bibliography{sample}

\IfFileExists{\jobname.bbl}{}
 {\typeout{}
  \typeout{******************************************}
  \typeout{** Please run "bibtex \jobname" to optain}
  \typeout{** the bibliography and then re-run LaTeX}
  \typeout{** twice to fix the references!}
  \typeout{******************************************}
  \typeout{}
 }


\end{document}